\documentclass[%
reprint, 
superscriptaddress,  
amsmath, 
amssymb, 
]{revtex4-2} %
\usepackage{graphicx}
\graphicspath{{../}}
\usepackage[utf8]{inputenc}
\usepackage[T1]{fontenc}
\usepackage{etoolbox}
\usepackage{mathptmx}
\usepackage{color}
\usepackage{soul}
\usepackage[section]{placeins}
\usepackage{tabularx}
\usepackage{array}
\usepackage{verbatim} 
\usepackage{xcolor}
\usepackage{bm}
\usepackage{hyperref}
\hypersetup{
   colorlinks=true,     
   linkcolor=blue,      
   citecolor=blue,      
   filecolor=blue,   
   urlcolor=blue       
}

\usepackage{siunitx}

\newcommand{\hc}{\ensuremath{\mathrm{H_{C}}}}

\newcommand{\roxy}{\ensuremath{\rho_{xy}}}
\newcommand{\roxx}{\ensuremath{\rho_{xx}}}
\DeclareSIUnit\angstrom{\text {Å}}
\DeclareSIUnit\bar{bar}

\begin{document}

\title{In-plane magnetic field control of anomalous Hall response enabled by magnetic anisotropy engineering}

\author{J. C. Rodriguez E.}
\affiliation{Instituto de Nanociencia y Nanotecnolog\'{i}a, CNEA--CONICET, Centro At\'{o}mico Bariloche, Av. E. Bustillo 9500 (R8402AGP), Bariloche, R\'{i}o Negro, Argentina.}
\affiliation{Instituto Balseiro, Universidad Nacional de Cuyo--CNEA, Av. E. Bustillo 9500 (R8402AGP) Bariloche, R\'{i}o Negro, Argentina.}\affiliation{Departamento Magnetismo y Materiales Magn\'{e}ticos, Gerencia de F\'{i}sica, Centro At\'{o}mico Bariloche, Av. E. Bustillo 9500 (R8402AGP) Bariloche, R\'{i}o Negro, Argentina.}

\author{E. De Biasi}
\affiliation{Instituto de Nanociencia y Nanotecnolog\'{i}a, CNEA--CONICET, Centro At\'{o}mico Bariloche, Av. E. Bustillo 9500 (R8402AGP), Bariloche, R\'{i}o Negro, Argentina.}
\affiliation{Instituto Balseiro, Universidad Nacional de Cuyo--CNEA, Av. E. Bustillo 9500 (R8402AGP) Bariloche, R\'{i}o Negro, Argentina.}\affiliation{Departamento Magnetismo y Materiales Magn\'{e}ticos, Gerencia de F\'{i}sica, Centro At\'{o}mico Bariloche, Av. E. Bustillo 9500 (R8402AGP) Bariloche, R\'{i}o Negro, Argentina.}

\author{J. I. Facio}
\affiliation{Instituto de Nanociencia y Nanotecnolog\'{i}a, CNEA--CONICET, Centro At\'{o}mico Bariloche, Av. E. Bustillo 9500 (R8402AGP), Bariloche, R\'{i}o Negro, Argentina.}
\affiliation{Instituto Balseiro, Universidad Nacional de Cuyo--CNEA, Av. E. Bustillo 9500 (R8402AGP) Bariloche, R\'{i}o Negro, Argentina.}%

\author{D. Salomoni}
\affiliation{Spintec, Universit\'{e} Grenoble Alpes, CNRS, CEA, Grenoble INP, IRIG-SPINTEC, 38000 Grenoble, France.}
\author{S. Auffret}
\affiliation{Spintec, Universit\'{e} Grenoble Alpes, CNRS, CEA, Grenoble INP, IRIG-SPINTEC, 38000 Grenoble, France.}
\author{R. C. Sousa}
\affiliation{Spintec, Universit\'{e} Grenoble Alpes, CNRS, CEA, Grenoble INP, IRIG-SPINTEC, 38000 Grenoble, France.}
\author{I. L. Prejbeanu}
\affiliation{Spintec, Universit\'{e} Grenoble Alpes, CNRS, CEA, Grenoble INP, IRIG-SPINTEC, 38000 Grenoble, France.}
\author{A. Bruchhausen}
\affiliation{Instituto de Nanociencia y Nanotecnolog\'{i}a, CNEA--CONICET, Centro At\'{o}mico Bariloche, Av. E. Bustillo 9500 (R8402AGP), Bariloche, R\'{i}o Negro, Argentina.}
\affiliation{Instituto Balseiro, Universidad Nacional de Cuyo--CNEA, Av. E. Bustillo 9500 (R8402AGP) Bariloche, R\'{i}o Negro, Argentina.}\affiliation{Departamento de F\'{o}tonica y Optoelectr\'{o}nica, Gerencia de F\'{i}sica, Centro At\'{o}mico Bariloche, Av. E. Bustillo 9500 (R8402AGP) Bariloche, R\'{i}o Negro, Argentina.}

\author{J. Curiale}
\affiliation{Instituto de Nanociencia y Nanotecnolog\'{i}a, CNEA--CONICET, Centro At\'{o}mico Bariloche, Av. E. Bustillo 9500 (R8402AGP), Bariloche, R\'{i}o Negro, Argentina.}
\affiliation{Instituto Balseiro, Universidad Nacional de Cuyo--CNEA, Av. E. Bustillo 9500 (R8402AGP) Bariloche, R\'{i}o Negro, Argentina.}\affiliation{Departamento Magnetismo y Materiales Magn\'{e}ticos, Gerencia de F\'{i}sica, Centro At\'{o}mico Bariloche, Av. E. Bustillo 9500 (R8402AGP) Bariloche, R\'{i}o Negro, Argentina.}

\author{L. Avilés-Félix}
\affiliation{Instituto de Nanociencia y Nanotecnolog\'{i}a, CNEA--CONICET, Centro At\'{o}mico Bariloche, Av. E. Bustillo 9500 (R8402AGP), Bariloche, R\'{i}o Negro, Argentina.}
\affiliation{Instituto Balseiro, Universidad Nacional de Cuyo--CNEA, Av. E. Bustillo 9500 (R8402AGP) Bariloche, R\'{i}o Negro, Argentina.}
\affiliation{Departamento Magnetismo y Materiales Magn\'{e}ticos, Gerencia de F\'{i}sica, Centro At\'{o}mico Bariloche, Av. E. Bustillo 9500 (R8402AGP) Bariloche, R\'{i}o Negro, Argentina.}
\date{\today}
\begin{abstract}
Engineering magnetic anisotropy provides a powerful route to control magnetization orientation and unlock emerging functionalities in opto-spintronic and current-driven devices. Beyond its role in magnetization reversal, the effective anisotropy can strongly influence the magnetotransport response, offering an additional degree of freedom to tune new device functionalities. In this work, we report a magnetotransport study of a ferrimagnetic [Tb/Co]$_{\times 5}$ multilayer grown with a Tb thickness gradient, whose wedge-shaped tilts the uniaxial anisotropy axis slightly away from the film normal. Anomalous Hall resistivity measurements from \qty{80}{\kelvin} to \qty{300}{\kelvin} reveal a spin reorientation transition, while the angular dependence of the magnetotransport responses exposes the crucial role of the tilted anisotropy.
A simplified macrospin model reproduces the full angular response across the transition and shows that the observed anomalous Hall effect when the in-plane magnetic field is applied originates from the tilt of the uniaxial anisotropy axis, which supplies a built-in symmetry-breaking mechanism, enabling in-plane field control over the out-of-plane anomalous Hall response, sign included. 
These findings establish tilted magnetic anisotropy as a promising route toward Hall effect-based sensor applications and highlight Tb/Co multilayers as a versatile platform for anisotropy-engineered spintronic devices.
\end{abstract}

\maketitle
The control of magnetization in ferrimagnetic rare-earth (RE)–transition-metal (TM) alloys and multilayers has attracted considerable attention owing to its potential for spintronic devices with novel functionalities. The ability to engineer and tune perpendicular magnetic anisotropy (PMA) is critical for applications such as ultrafast all-optical switching at the nanoscale \cite{Mondal2023,Mishra2023,Olivier2020, WZhang2025}, spin valves, magnetic tunnel junctions\cite{Wu2020} and current-induced spin–orbit torque devices for nonvolatile memory technologies~\cite{Wong2019,Kim2022b,Hu2024,Liu2024}. More broadly, controlling the orientation of the magnetization rather than merely its magnitude underlies a range of opto-spintronic and current-driven functionalities.

Among ferrimagnetic multilayers, Tb/Co heterostructures are particularly attractive because they combine a strong PMA with antiferromagnetic exchange coupling between the RE and TM sublattices; their magnetic properties can be tailored by a number of handles, such as layer thickness, number of repetitions, interface structure, and temperature, thereby providing a versatile platform for engineering the magnetic state. Crucially, the tilt of the uniaxial anisotropy axis away from the film normal -- usually an incidental consequence of off-axis growth~\cite{JRodriguezJMMM2026,SalomoniAPS2023} -- can instead serve as a built-in structural resource: by lifting the up--down ($z\to-z$) symmetry, it endows the magnetization with a finite out-of-plane component even under a nominally in-plane magnetic field. Importantly, this component depends on the field orientation, which can be modulated and even reversed by an in-plane field rotation. 

Such built-in symmetry breaking raises the prospect that a purely in-plane field could control transport signatures normally associated with the out-of-plane magnetization ($M_z$), a possibility that, to our knowledge, remains largely unexplored in ferrimagnetic multilayers. This phenomenology differs from the anomalous in-plane Hall effect recently reported in Fe$_3$Sn, where the in-plane field directly modulates the Hall conductivity through Zeeman coupling in a spin-canted magnetic state, rather than acting primarily through a field-induced variation of $M_z$~\cite{sankar2025room}.

Magnetotransport  provides a sensitive probe of this behavior. The anomalous Hall effect (AHE) provides direct access to the out-of-plane magnetization and is routinely used to determine the coercive field, its temperature dependence, and changes in the magnetic configuration, while the planar Hall effect (PHE) is sensitive to the in-plane magnetization component. The combination of these responses is especially valuable when the magnetization is neither confined to the film plane nor to the film normal, which is precisely the situation created by a tilted anisotropy axis, where AHE and PHE contribute simultaneously to the transverse signal. 

\begin{figure*}[ht]
                \includegraphics[width=\textwidth]{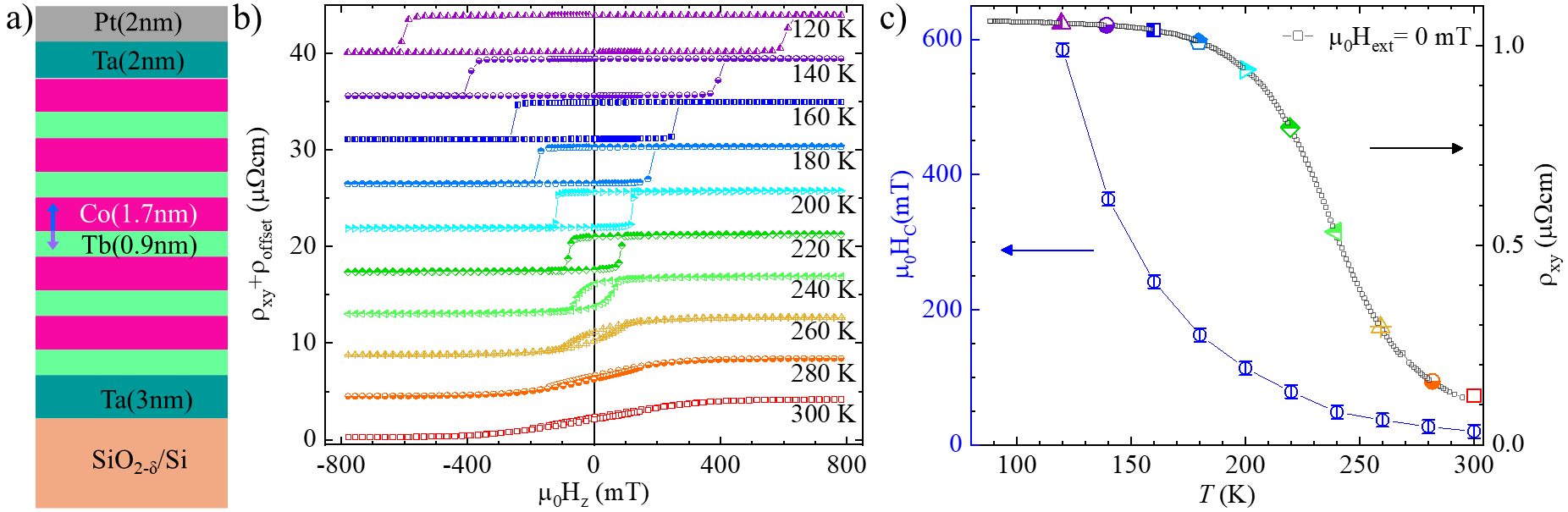}
                \caption{\label{fig:SampleStructure}
                 \textbf{Sample structure and magneto-transport response in Tb/Co multilayers}. a) Schematic of TbCo(0.9) ferrimagnetic multilayers. b) The temperature-evolution of the transverse (\roxy) resistivity as a function of magnetic field in out-of-plane (OOP) field sweep for the TbCo(0.9) multilayers Hall bar. c) The temperature dependence of the transverse (\roxy) resistivity (black open squares) and the coercive fields (blue open circles) extracted from the different measurement shown in panel b) for the same sample. The \roxy~for each temperature shown in panel b) is also highlighted.
                } 
\end{figure*} 

In this work, we report AHE in Tb/Co multilayers under in-plane magnetic field rotation, persisting close to room temperature. Combining the angular dependence of magnetotransport with a harmonic analysis, we trace this signal to a small tilt of the uniaxial anisotropy axis, which drives a finite out-of-plane magnetization component whose sign, and hence that of the AHE, is set by the in-plane field direction. A simplified macrospin model, incorporating the temperature dependence of the saturation magnetization and the uniaxial anisotropy, reproduces the full angular response across the temperature-driven spin-reorientation transition, from the easy-axis to the easy-plane configuration.


We grew Tb/Co ferrimagnetic multilayers on a 4-inch thermally-oxidized silicon wafer (SiO$_2\approx$ \qty{500}{\nm}) in Co-rich composition by DC magnetron sputtering\cite{JRodriguezJMMM2026}. Particularly, the multilayer structure of the sample studied consist of Ta(3)/[Tb(0.9)/Co(1.7))]$_{\times 5}$/Ta(2)/Pt(2) as schematized in Fig.~\ref{fig:SampleStructure}a). The multilayer film was lithographed into \qty{10}{\um}~$\times$~\qty{900}{\um} Hall bar for measurements of longitudinal resistivity (\roxx) and transverse resistivity (\roxy) simultaneously. The thickness of each layer is indicated in parenthesis in \qty{}{\nm}.

The magnetotransport measurements were performed with a low in-plane DC current of \qty{1}{\mA} using the home-made setup described in Ref. \cite{Saba_2025}. The $\hat{x}$ axis is defined by the sense of the applied current $I$. The sample was mounted inside a Janis SVT-300 cryostat, which allowed us to control the temperature in the range of \qty{77} - \qty{300}{\kelvin}. The magnetic field could be continuously rotated with respect to the sample, allowing measurements over a range of angles relative to the film plane.


Fig.~\ref{fig:SampleStructure}b) shows the \roxy~loops measured for the Tb/Co sample at various temperatures, with the magnetic film applied perpendicular to the plane ($\mathrm{H_{z}}$). At low temperatures from \qty{120} to \qty{220}{\kelvin}, square hysteresis loops, typically observed in systems with strong PMA, were obtained. In contrast, the loops deviate from this square-shaped at temperatures between \qty{240} to \qty{300}{\kelvin}. The values of the coercive field (\hc) obtained within the \qty{120} - \qty{300}{\kelvin} range are plotted in Fig.~\ref{fig:SampleStructure}c). Upon cooling, the \hc~increases because the magnetic moment of the Tb sub-lattice increases, tending to be comparable with the Co sub-lattice, as previously reported \cite{JRodriguezJMMM2026}. The AHE loops of $\mathrm{Tb}_{1-x}\mathrm{Co}_{x}$ alloys in Co-rich sample show a similar trend of the \hc\cite{HeAIP2026,UedaPRB2017,PhamPRA2018}. In addition, the loop polarity remains unchanged across the entire temperature range, as shown in Fig.~\ref{fig:SampleStructure}b). This behavior indicates that the anomalous Hall signal is dominated by the TM sublattice (Co) within the measured temperature window, as typically observed in Co-rich samples \cite{HeAIP2026}.

Conversely, the almost linear dependence of the \roxy~as a function of the magnetic field $\mathrm{H_{z}}$  suggests that the $\hat{z}$ axis is a hard axis for the temperature range \qty{240} - \qty{300}{\kelvin}. The contribution from the planar Hall effect (PHE) is very small in the same temperature range. These results indicate a rotation of the easy axis (also called a spin-reorientation transition) as a function of temperature in the Tb/Co multilayer Hall bar. This phenomenon of the rotation of the easy axis was reported in Tb/Co multilayers as a function of temperature and Tb thickness \cite{GarreauPRB1996} which can be explained by the competition between the shape and uniaxial anisotropies.

\begin{figure*}[htbp]
                \includegraphics[width=\textwidth]{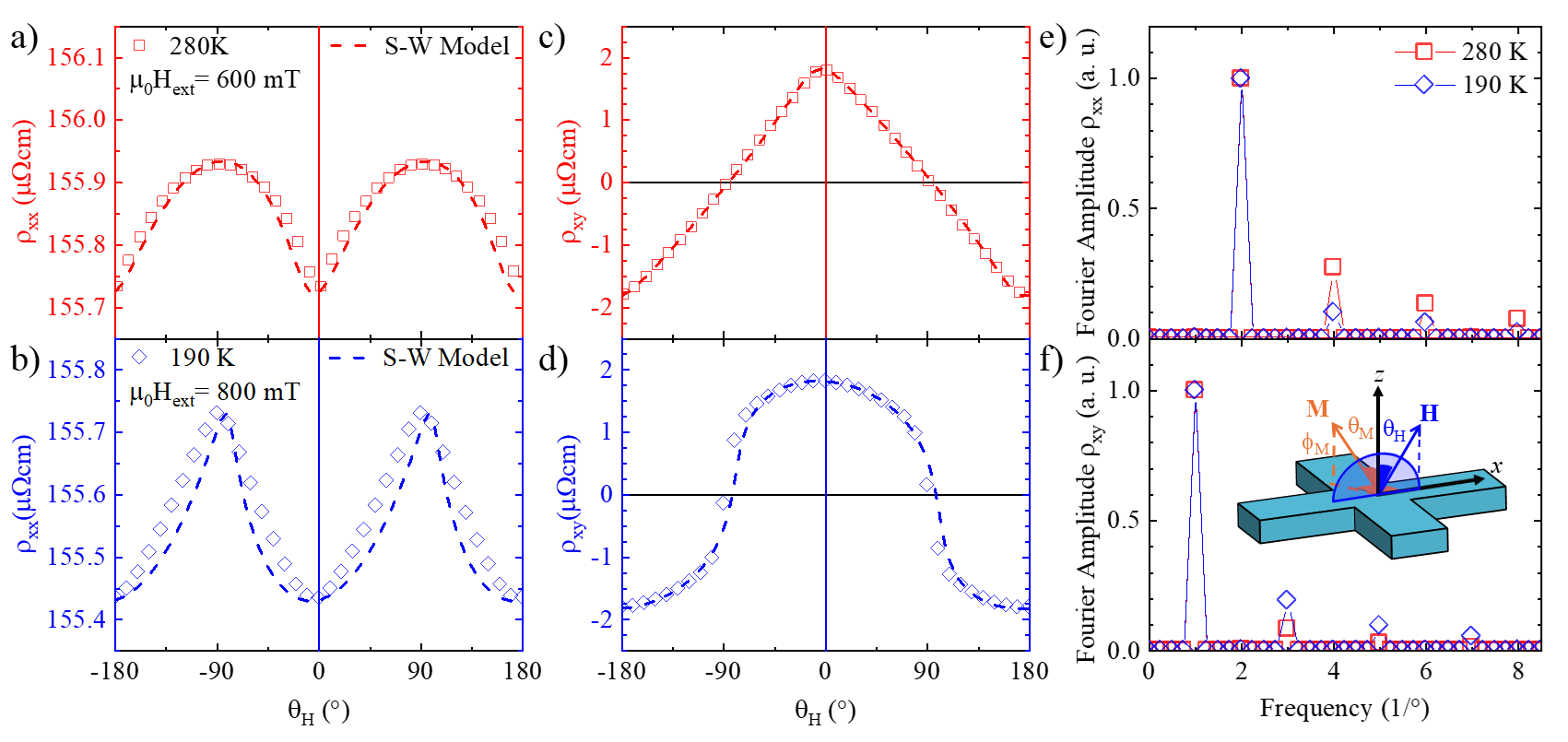}
                \caption{\label{fig:RvTheta}
                \textbf{Longitudinal (\roxx) and transverse (\roxy) resistivities as function of the polar angle of the external magnetic field ($\theta_H$)}. Panels a) and c) show the longitudinal and transverse resistivities, respectively, as a function of the $\theta_H$ measured at \qty{280}{\kelvin} and \qty{600}{\mT}. Panels b) and d) show the corresponding resistivities measured at \qty{190}{\kelvin} and \qty{800}{\mT}, respectively. 
                Symbols represent the experimental data, while dashed lines correspond to simulations based on the S–W model. e) Normalized Fourier amplitude of the angular dependence of \roxx~acquired at \qty{280}{\kelvin} and \qty{190}{\kelvin} shown in panels a) and c). f) Normalized Fourier amplitude of the angular dependence of \roxy~acquired at \qty{280}{\kelvin} and \qty{190}{\kelvin} shown in panels b) and d). The Fourier amplitude showed in panels e) and f) were normalized to frequency with the highest amplitude.
                The inset in panel f) shows the schematic representation of OOP geometry rotation of the magnetic field.
}
\end{figure*}

To obtain more evidence regarding the rotation of the easy axis, we measured the \roxy~as a function of the temperature in the remanent state for the Tb/Co sample, that is, after applying a magnetic field perpendicular to the plane and saturating it at \qty{300}{\kelvin}, as shown in Fig.~\ref{fig:SampleStructure}c). It is evidenced that the \roxy~increases monotonically as the sample is cooled down and reaches a plateau at low temperature, whereas the \roxx~remains nearly constant within the same temperature range. The resistivity $\rho_{\mathrm{Tb/Co}}$ is $\sim156~\mu\Omega$\qty{}{\cm} in consistent with previous reports in alloys of similar composition\footnote{Same composition corresponds to multilayers where the normalized amount of each metal is equivalent to the alloy's atomic fraction $\mathrm{Tb}_{x}\mathrm{Co}_{1-x}$} \cite{KimJAP2000, PhamPRA2018}. All these results confirm a continuous rotation of the magnetization from out-of-plane (OOP) to in-plane (IP) as the temperature increases.

We next discuss angle-dependent magnetoresistance measurements performed on the Tb/Co sample for different polar and azimuthal angles. The measurements were acquired under applied external magnetic fields of \qty{600}{\mT} at \qty{280}{\kelvin} and \qty{800}{\mT} at \qty{190}{\kelvin}.
As shown in Fig. \ref{fig:SampleStructure}b), the sample exhibits dominant PMA at \qty{190}{\kelvin}, while the in-plane anisotropy becomes dominant at \qty{280}{\kelvin}.

We first performed the angular dependence of the longitudinal and transverse resistivity by subjecting the Tb/Co sample to a rotational field in the $\hat{x}$-$\hat{z}$ plane, see Figs.~\ref{fig:RvTheta}a)-d). The polar angle $\theta_H$ denotes the angle between the applied field and $\hat{z}$ axis when the field rotates from $\hat{z}$-axis to $\hat{x}$-axis in the $\hat{x}$-$\hat{z}$ plane as shown in the inset in Fig.~\ref{fig:RvTheta}f).

Figs.~\ref{fig:RvTheta}a) and b) show the $\theta_H$ dependence of the \roxx~measured at \qty{190}{\kelvin} and \qty{280}{\kelvin} and under an external magnetic field of \qty{600}{\mT} and \qty{800}{\mT}, respectively. For this field rotation geometry, the longitudinal resistivity is expected to be dominated by the anisotropy magnetoresistance (AMR), as typically observed in ferromagnets. However, the \roxx~clearly deviates from the sine-squared dependence ($\sin^{2}\theta_{H}$). This behavior indicates that the magnetization does not continuously follow the direction of the applied magnetic field during the rotation. This deviation is further supported by the Fourier analysis of the \roxx, shown in Fig.~\ref{fig:RvTheta}e), which reveals a dominant second harmonic together with higher-order even harmonic. The presence of these additional harmonic components demonstrates that the longitudinal magnetotransport response cannot be described solely by the two-fold AMR symmetry. In fact, these higher-order harmonics reflect the influence of the magnetic anisotropy, where the magnetization prefers to point, on the resistivity. For instance, a notable feature is observed at $\theta_H = 0^{\circ}$ and $90^{\circ}$. At \qty{190}{\kelvin}, the \roxx~exhibits a sharp peak at $\theta_H = 90^{\circ}$ and a broad peak at $\theta_H = 0^{\circ}$, whereas the opposite behavior is observed at \qty{280}{\kelvin}, as shown in Fig.~\ref{fig:RvTheta}a) and b). The distinct angular dependence of \roxx~observed at \qty{190}{\kelvin} and \qty{280}{\kelvin} reflects the temperature-driven spin reorientation transition from easy axis to an easy plane magnetic configuration, consistent with the results presented in Fig.~\ref{fig:SampleStructure}b) and c).

In addition, we measured the transverse resistivity \roxy~under the same conditions described above. The \roxy~values were obtained from the raw Hall resistivity \roxy~after subtracting the contribution arising from the misalignment of the Hall contacts. In general, the \roxy~is dominated by AHE for the OOP geometry because the ordinary Hall effect contribution is quite small, as usually observed in ferrimagnet \cite{KimJAP2000}. Figs.~\ref{fig:RvTheta}c) and d) show the $\theta_{H}$ dependence of the \roxy~measured at \qty{280}{\kelvin} and \qty{190}{\kelvin}, respectively. As shown in the Fig.~\ref{fig:RvTheta}d), the angular variation of \roxy~exhibits a nearly square-wave curve with one-fold symmetry under OOP geometry rotation, consistent with a dominant PMA. Accordingly, the $\theta_{H}$  dependence of the \roxy~indicates a switching of the magnetization between two well-defined levels, corresponding to an abrupt reversal of $M_{z}$ during the field rotation around $\theta_H = 90^\circ$.

In contrast, in Fig.~\ref{fig:RvTheta}c), the angular variation of $\roxy$ displays a almost triangular-wave curve with one-fold symmetry at \qty{280}{\kelvin}. That is, the \roxy~ reaches a a maximum at $\theta_H = 0^\circ$ and decreases nearly linearly as the magnetic field is rotated toward the opposite direction. This behavior can be explained by the magnetization lying predominantly in the plane at this temperature. As the magnetic field is rotated, it gradually pulls the magnetization out of the plane, leading to a continuous increase in $M_{z}$ and reaching maximum when the magnetization is perpendicular to the film plane $\theta_H = 0^{\circ}$.

Consequently, the AHE resistivity shown in Fig.~\ref{fig:RvTheta}c) and d) are consistent with the observation that the magnetization vector is not aligned with the external field at either temperature. This deviation between the magnetization and the applied field is further supported by the Fourier analysis of the \roxy~, shown in Fig.~\ref{fig:RvTheta}f), which exhibits only odd harmonics for both measurements. One notable feature of the triangle curve is that its higher-order harmonics decay much faster than those of a square curve, scaling as the inverse square of the harmonic number, in contrast to the inverse scaling observed for the square curve, as seen in Fig.~\ref{fig:RvTheta}f). Finally, no significant contribution from a PHE, which would exhibit a two-fold symmetry, is observed for this field rotation at either temperature.

Since the magnetization is not aligned with the external field, determining the contributions of AMR and AHE from the angular field rotation measurements is not straightforward. Furthermore, the mechanism responsible for the rotation of the easy axis observed in Fig.~\ref{fig:SampleStructure} and Fig.~\ref{fig:RvTheta} has not yet been clarified. In this scenario, we performed numerical simulations based on the Stoner-Wolfarth (S-W) formalism \cite{JRodriguezJMMM2026}, assuming in which the magnetic behavior of the system is approximated by a single macrospin ($\overrightarrow{M} =  M_{s} \hat{m}$) with only 2 degrees of freedom, $\theta_{M}$ and $\phi_{M}$. Therefore, the angular dependence of resistivity was obtained using the generalized Ohm’s law \cite{Campbell1982,Harder2016}. The contributions from the ordinary magnetoresistance and the ordinary Hall effect were neglected in this system. 

The parameters used in the simulation of the $\theta_{H}$ dependence of the \roxy~and \roxx~were set to $\mathrm{M_S}=\qty{600}{\kA/m}$ for \qty{280}{\kelvin}. The value of $M_s$ used in the simulation is consistent with the values reported for samples in which the magnetization lies into the plane  \cite{Alebrand2012,Thorarinsdottir2023}. A tilt of uniaxial anisotropy axis $\hat{n}$ with respect to the out-of-plane direction is imposed to account for the off-axis film growth \cite{JRodriguezJMMM2026, SalomoniAPS2023}.

The simulated \roxx~and \roxy~plotted as dashed lines in the Fig.~\ref{fig:RvTheta}a)-d) show a good agreement with the experimental data. The best results were obtained for a deviation of the anisotropy axis $\hat{\eta}$ from $\hat{z}$ into the $\hat{z}$-$\hat{x}$ plane of $\theta_{K}=5^{\circ}$ at both temperatures, consistent with previous reports \cite{JRodriguezJMMM2026, SalomoniAPS2023}. 

\begin{figure*}[htbp]
            \includegraphics[width=\textwidth]{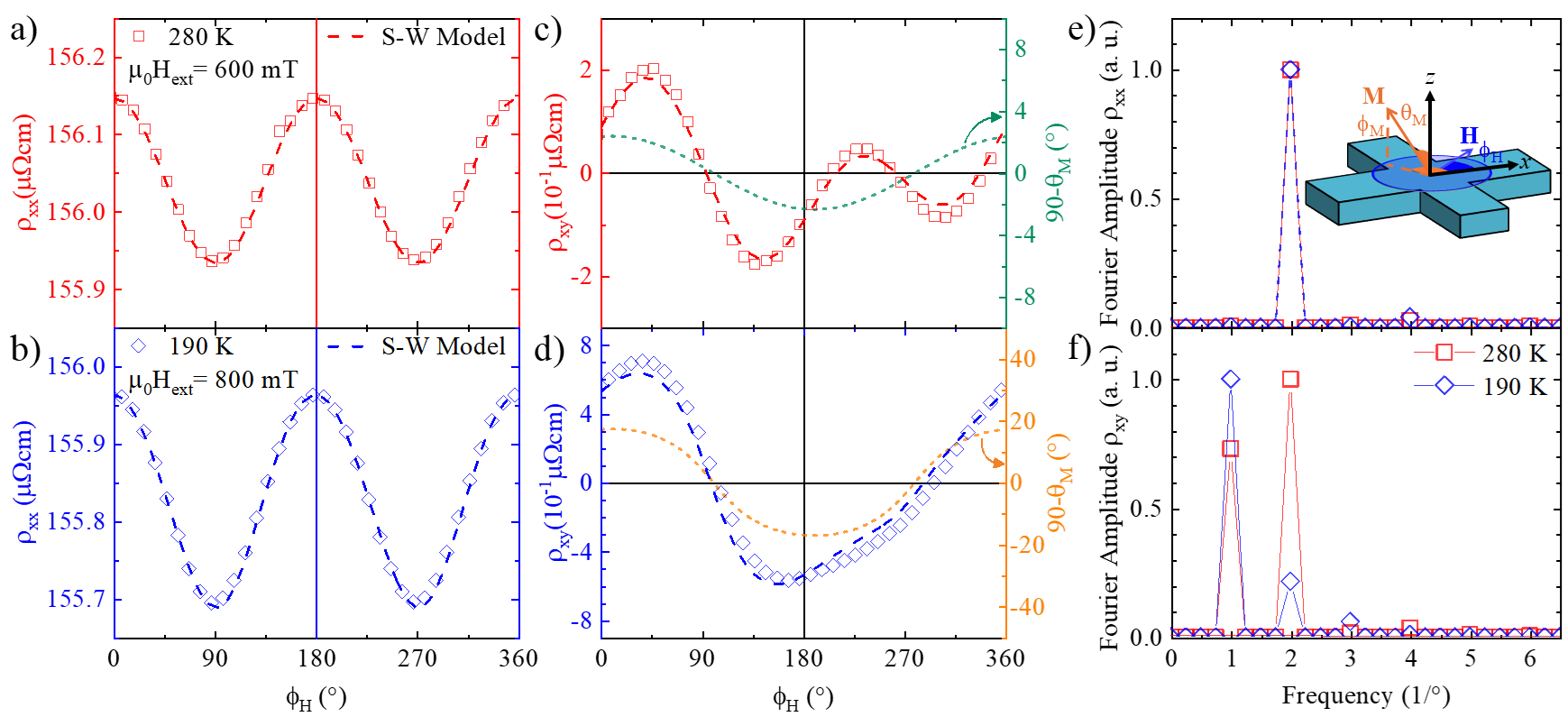}            
            \caption{\label{fig:RvPhi}
            \textbf{Longitudinal (\roxx) and transverse (\roxy) resistivity as a function of in-plane field orientation}.
            Panels a) and c) show the longitudinal and transverse resistivities, respectively, as a function of the $\theta_H$ measured at \qty{280}{\kelvin} and \qty{600}{\mT}. Panels b) and d) show the corresponding resistivities measured at \qty{190}{\kelvin} and \qty{800}{\mT}, respectively. 
            Symbols represent the experimental data, while dashed lines correspond to simulations based on the S–W model.
            The short dash lines shown in panels c) and d) correspond to the deviation of the the equilibrium polar angle $\theta_{M}$ from the equatorial plane obtained by the S-W model as the in-plane magnetic field rotates.
            e) Normalized Fourier amplitude of the \roxx~resistivities from the angular variation shown in panels a) and c). f) Normalized Fourier amplitude of the \roxy~resistivities from the angular variation shown in panels b) y d). The Fourier amplitude showed in panels e) y f) were normalized to frequency with the highest amplitude.
            The inset in panel e) shows the schematic representation of IP geometry rotation of the magnetic field.
}
\end{figure*}

The simulations also suggest a relative angle between the projection of the anisotropy axis into the $\hat{x}$-$\hat{y}$ plane and $\hat{x}$ axis of $\phi_{K}=10^{\circ}$. Moreover, the macrospin model indicates that the shape anisotropy $\textrm{K}_\textrm{s}$ exceeds the uniaxial anisotropy $\textrm{K}_\textrm{u}$ at \qty{280}{\kelvin}, resulting in an easy plane. In contrast, at \qty{190}{\kelvin}, the $\textrm{K}_\textrm{u}$ dominates over $\textrm{K}_\textrm{s}$, leading to an easy axis. It is important to note that the direction of the $\textrm{K}_\textrm{u}$ remains the same in both cases.

We also investigated the angular dependence of the longitudinal and transverse resistivity by subjecting the sample to a rotational field in the $\hat{x}$-$\hat{y}$ plane, as shown in Fig.~\ref{fig:RvPhi}a)-d). The azimuthal angle $\phi_{H}$ is defined as the angle between the rotating field and $\hat{x}$ axis when the field rotates in the $\hat{x}$-$\hat{y}$ plane, as shown in the inset in Fig.~\ref{fig:RvPhi}e).

Fig.~\ref{fig:RvPhi}a) and b) show the $\phi_{H}$  dependence of \roxx~measured at \qty{190}{\kelvin} and \qty{280}{\kelvin} and under an external magnetic field of \qty{600}{\mT} and \qty{800}{\mT}, respectively. In contrast with the OOP rotation variation, both measurements of \roxx~follow a nearly cosine-squared dependence ($\cos^{2}\phi_{H}$) with a clear twofold symmetry, as expected for AMR. In addition, the AMR amplitude is slightly larger at \qty{190}{\kelvin} than at \qty{280}{\kelvin}. The Fourier analysis of \roxx, shown in Fig. \ref{fig:RvPhi}e), reveals a dominant second harmonic, further confirming the AMR origin of the signal with AMR values of $\sim 0.3 \%$. These results indicate that the magnetization closely follows the direction of the magnetic field, even at \qty{190}{\kelvin} when the easy axis is almost out of the film plane. The simulated \roxx, plotted as dashed lines in Figs. \ref{fig:RvPhi}a) and b), shows excellent agreement with the experimental data. Notably, these simulations were obtained using the same set of parameters employed for the OOP rotation simulations. A remarkable feature of this geometry is that the longitudinal resistivity is essentially insensitive to the tilt of the uniaxial anisotropy axis, making the $\phi_{H}$ angular dependence of \roxx~primarily determined by the AMR contribution.

We also measured the $\phi_{H}$ dependence of the \roxy~at \qty{190}{\kelvin} and \qty{280}{\kelvin}, as shown in Figs.~\ref{fig:RvPhi}c) and d), respectively. At \qty{280}{\kelvin}, the \roxy~exhibits a nearly harmonic angular dependence, whereas at \qty{190}{\kelvin}, it displays an approximately square-wave-like profile. Both measurements deviate significantly from the two-fold symmetry expected for the PHE. The Fourier analysis of the \roxy, shown in Fig.~\ref{fig:RvPhi}f), reveals a dominant second harmonic together with an unexpected first harmonic  at \qty{280}{\kelvin}. Notably, at \qty{190}{\kelvin} the amplitude of the first harmonic exceeds that of the second harmonic. The presence of the first harmonic component suggests a contribution from the AHE, indicating that the magnetization is not entirely on the film plane.

To further understand the $\phi_{H}$ angular dependence of the \roxy, we performed simulations of the transverse resistivity, shown as dashed lines in Fig.~\ref{fig:RvPhi}c) and d), using the same set of parameters. The good agreement between the experimental data and simulations indicates that the observed AHE contribution during in-plane field rotation originates from the tilt of the uniaxial anisotropy axis. In particular, a tilt of $\theta_{K}=5^{\circ}$ from the $\hat{z}$ axis gives rise to a finite component $M_{z}$, resulting in the appearance of a AHE signal in the in-plane rotation of the magnetic field.

The deviation of the polar angle $\theta_{M}$ from the equatorial plane obtained by numerical simulations during the in-plane field rotation are shown as short dashed lines at \qty{280}{\kelvin} and \qty{190}{\kelvin} in the Fig.~\ref{fig:RvPhi}c) and d), respectively.
The simulations show that the magnetization does not remain strictly confined to the film plane, but instead oscillates about the equatorial plane as the magnetic field rotates. At \qty{280}{\kelvin} , the maximum deviation from the equatorial plane is approximately $\pm 3^{\circ}$, occurring near the $\hat{x}$ axis. At \qty{190}{\kelvin} , this deviation increases to approximately $\pm 17^{\circ}$, indicating the stronger influence of the tilted anisotropy axis on the equilibrium magnetization direction.

Therefore, we demonstrate that the AHE observed when the in-plane field is applied in Tb/Co multilayers originates from the tilt of the anisotropy axis induced by the off-axis growth of the film \cite{JRodriguezJMMM2026}. In addition to providing insights into the influence of tilted magnetic anisotropy on magnetotransport, this unique feature, which is difficult to realize through conventional approaches, may enable new functionalities in spintronic devices and Hall sensor applications. For instance, deterministic spin-orbit torque (SOT) switching typically requires an external in-plane magnetic field to break the symmetry and achieve controlled magnetization reversal \cite{PhamPRA2018, Wong2019, Liu2024}. In contrast, the built-in anisotropy tilt in Tb/Co Hall bar devices could provide the necessary symmetry breaking, potentially enabling field-free SOT switching.

On the other hand, enhancing the AHE signal when in-plane field is applied is attractive and useful for Hall sensor applications. As shown in Fig.~\ref{fig:RvPhi}c), the maximum deviation of the $\theta_{M}$ from the film plane at \qty{280}{\kelvin} is approximately $3^\circ$, corresponding to $m_{z} \sim 0.05$. Several strategies can be envisioned to overcome this limitation. First, the magnetic properties of ferrimagnetic materials can be highly tuned by temperature or composition. Consequently, the AHE signal can also be tuned by these parameters, as demonstrated in Fig.~\ref{fig:RvPhi}d). At \qty{190}{\kelvin}, the deviation from the film plane increases to approximately $17^\circ$, corresponding to $m_{z} \sim 0.3$. Consequently, the AHE signal increases by nearly a factor of five. A similar enhancement could be achieved by reducing the external magnetic field or increasing the Tb layer thickness, both of which favor a larger $M_{z}$. Second, the simulation reveals a relative angle of $\phi_{K} = 10^\circ$ between the projection of the anisotropy axis and the Hall bar ($\hat{x}$ axis). According to numerical simulations, increasing the $\phi_{K}$, for example to $45^\circ$ through device design and lithographic patterning, is expected to significantly improve the transverse resistivity signal. In this case, the PHE and AHE contributions become nearly in phase, leading to a substantially larger signal amplitude.


In summary, we demonstrated that an in-plane magnetic field can control the observed out-of-plane AHE in [Tb(\qty{1.9})/Co(\qty{1.7}))]$_{\times 5}$ multilayers. The magnetotransport properties were systematically investigated for two representative magnetic configurations, corresponding to an easy-axis state at \qty{190}{\kelvin} and an easy-plane state at \qty{280}{\kelvin}, revealing how the tilted anisotropy governs the angular dependence of both the anomalous and planar Hall responses. The resistivity loops acquired at different temperatures further demonstrate a continuous evolution of the nearly perpendicular effective anisotropy axis, from an easy-axis at low temperature to a hard-axis at high temperature. The angular dependence of both the longitudinal and transverse resistivity are in good agreement with calculations based on a simplified macrospin model. Beyond its fundamental interest, the tilted magnetic anisotropy and appearance of AHE when in-plane field is applied may provide a route toward field-free SOT switching and improved Hall sensor functionalities.

\section*{Acknowledgements}
Technical support from Rubén E. Benavides, César Pérez, Matías Guillén and Miguel Rubio Roy is greatly acknowledged. This work was partially supported by Conicet under Grant PIBAA 2022-2023 (project MAGNETS) Grant ID. 28720210100099CO; ANPCyT Grant PICT-2018-01138, ANPCyT Grant PICT 2021-00113 (project DISCO) and U.N. Cuyo Grant 06/C556 from Argentina. This work was supported by the French ANR under projects SPINMAT (PEPR SPIN ANR-22-EXSP-0007), SPINCHARAC (PEPR SPIN ANR-22-EXSP-0006) and NANOFUTUR (ESR/EQUIPEX+ ANR-21-ESRE-0012). We also acknowledge the financial support of European Commission by the H2020-MSCA RISE project ULTIMATE-I (Grant No. 101007825). L. Avilés-Félix thanks TRINH Quoc-Trung for fruitful discussions, Claudio Ferrari from the Departamento de Micro y Nanotecnología (CNEA - INN) for the preparation of the lithography masks and S. Anguiano and Diego Pérez for their assistance with the microfabrication facilities.

\bibliography{TbCoTransporte}

@article{sankar2025room,
  title={{Room temperature observation of the anomalous in-plane Hall effect in a Weyl ferromagnet}},
  author={Sankar, Soumya and Cheng, Xingkai and Murtaza, Tahir and Chen, Caiyun and Qin, Yuqi and Wu, Xuezhao and Shao, Qiming and Lortz, Rolf and Liu, Junwei and J{\"a}ck, Berthold},
  journal={Nat. Commun.},
  volume={17},
  number={1},
  pages={423},
  year={2025},
  url={https://www.nature.com/articles/s41467-025-67111-x},
  publisher={Nature Publishing Group UK London}
}

@article{JRodriguezJMMM2026,
    title = {Perpendicularly magnetized Tb/Co multilayers featuring tilted uniaxial anisotropy: Experiments and modeling},
    journal = {Journal of Magnetism and Magnetic Materials},
    volume = {642},
    pages = {173830},
    year = {2026},
    issn = {0304-8853},
    doi = {https://doi.org/10.1016/j.jmmm.2026.173830},
    url = {https://www.sciencedirect.com/science/article/pii/S0304885326000211},
    author = {J.C. {Rodriguez E.} and L. Avilés-Félix and M.H. Aguirre and L.M. Rodríguez and D. Salomoni and S. Auffret and R.C. Sousa and I.L. Prejbeanu and A.E. Bruchhausen and E. {De Biasi} and J. Curiale}
    }

@article{Saba_2025,
    doi = {10.1088/1361-6463/adf11b},
    url = {https://doi.org/10.1088/1361-6463/adf11b},
    year = {2025},
    month = {jul},
    publisher = {IOP Publishing},
    volume = {58},
    number = {30},
    pages = {305005},
    author = {Saba, L and Gómez, J E and Pérez-Morelo, D J and Anguiano, S and Velázquez Rodriguez, D and Butera, A and Granada, M and Avilés-Félix, L},
    title = {Magnetization process in epitaxial \mathrm{Fe}_{85}\mathrm{Co}_{15} thin films via anisotropic magnetoresistance},
    journal = {Journal of Physics D: Applied Physics}
    }

@article{GarreauPRB1996,
    title = {Spin-reorientation transition in ultrathin Tb/Co films},
    author = {Garreau, G. and Beaurepaire, E. and Ounadjela, K. and Farle, M.},
    journal = {Phys. Rev. B},
    volume = {53},
    issue = {3},
    pages = {1083--1086},
    numpages = {0},
    year = {1996},
    month = {Jan},
    publisher = {American Physical Society},
    doi = {10.1103/PhysRevB.53.1083},
    url = {https://link.aps.org/doi/10.1103/PhysRevB.53.1083}
    }

@article{HeAIP2026,
    author = {He, Tao and Dai, Cailan and Zhu, Hongtao and Zhou, Yu and He, Xiong and Xu, Yunli and Pan, Liqing and Lu, Guangduo},
    title = {Composition and temperature dependent magnetic transport and magnetization switching in W/Tb1−xCox/W multilayer films},
    journal = {AIP Advances},
    volume = {16},
    number = {5},
    pages = {055205},
    year = {2026},
    month = {05},
    issn = {2158-3226},
    doi = {10.1063/5.0333535},
    url = {https://doi.org/10.1063/5.0333535},
    }

@article{SalomoniAPS2023,
	author = {Salomoni, D. and Peng, Y. and Farcis, L. and Auffret, S. and Hehn, M. and Malinowski, G. and Mangin, S. and Dieny, B. and Buda-Prejbeanu, L.D. and Sousa, R.C. and Prejbeanu, I.L.},
	doi = {10.1103/PhysRevApplied.20.034070},
	issue = {3},
	journal = {Phys. Rev. Appl.},
	month = {Sep},
	numpages = {8},
	pages = {034070},
	publisher = {American Physical Society},
	title = {Field-Free All-Optical Switching and Electrical Readout of $\mathrm{Tb}$/$\mathrm{Co}$-Based Magnetic Tunnel Junctions},
	url = {https://link.aps.org/doi/10.1103/PhysRevApplied.20.034070},
	volume = {20},
	year = {2023}
    }

@article{UedaPRB2017,
    title = {Temperature dependence of spin-orbit torques across the magnetic compensation point in a ferrimagnetic TbCo alloy film},
    author = {Ueda, Kohei and Mann, Maxwell and de Brouwer, Paul W. P. and Bono, David and Beach, Geoffrey S. D.},
    journal = {Phys. Rev. B},
    volume = {96},
    issue = {6},
    pages = {064410},
    numpages = {6},
    year = {2017},
    month = {Aug},
    publisher = {American Physical Society},
    doi = {10.1103/PhysRevB.96.064410},
    url = {https://link.aps.org/doi/10.1103/PhysRevB.96.064410}
    }

@article{PhamPRA2018,
    title = {Thermal Contribution to the Spin-Orbit Torque in Metallic-Ferrimagnetic Systems},
    author = {Pham, Thai Ha and Je, S.-G. and Vallobra, P. and Fache, T. and Lacour, D. and Malinowski, G. and Cyrille, M. C. and Gaudin, G. and Boulle, O. and Hehn, M. and Rojas-S\'anchez, J.-C. and Mangin, S.},
    journal = {Phys. Rev. Appl.},
    volume = {9},
    issue = {6},
    pages = {064032},
    numpages = {9},
    year = {2018},
    month = {Jun},
    publisher = {American Physical Society},
    doi = {10.1103/PhysRevApplied.9.064032},
    url = {https://link.aps.org/doi/10.1103/PhysRevApplied.9.064032}
}

@article{KimJAP2000,
    author = {Kim, T. W. and Gambino, R. J.},
    title = {Composition dependence of the Hall effect in amorphous TbxCo1−x thin films},
    journal = {Journal of Applied Physics},
    volume = {87},
    number = {4},
    pages = {1869-1873},
    year = {2000},
    month = {02},
    issn = {0021-8979},
    doi = {10.1063/1.372106},
    url = {https://doi.org/10.1063/1.372106},
}

@incollection{Campbell1982,
    title = {Chapter 9 Transport properties of ferromagnets},
    series = {Handbook of Ferromagnetic Materials},
    publisher = {Elsevier},
    volume = {3},
    pages = {747-804},
    year = {1982},
    issn = {1574-9304},
    doi = {https://doi.org/10.1016/S1574-9304(05)80095-1},
    url = {https://www.sciencedirect.com/science/article/pii/S1574930405800951},
    author = {I.A. Campbell and A. Fert}
}

@article{Thorarinsdottir2023,
    year = {2023},
    month = {mar},
    publisher = {IOP Publishing},
    volume = {35},
    number = {20},
    pages = {205802},
    author = {K A Thórarinsdóttir and B R Thorbjarnardóttir and U B Arnalds and F Magnus},
    title = {Competing interface and bulk anisotropies in Co-rich TbCo amorphous thin films},
    journal = {Journal of Physics: Condensed Matter},
}

@article{Harder2016,
    title = {Electrical detection of magnetization dynamics via spin rectification effects},
    journal = {Physics Reports},
    volume = {661},
    pages = {1-59},
    year = {2016},
    note = {Electrical detection of magnetization dynamics via spin rectification effects},
    issn = {0370-1573},
    doi = {https://doi.org/10.1016/j.physrep.2016.10.002},
    url = {https://www.sciencedirect.com/science/article/pii/S0370157316303167},
    author = {Michael Harder and Yongsheng Gui and Can-Ming Hu}
}

@article{WZhang2025,
author = {Zhang, Wei and Hehn, Michel and Peng, Yi and Gorchon, Jon and Remy, Quentin and Lin, Jun Xiao and Hohlfeld, Julius and Malinowski, Grégory and Zhao, Wei Sheng and Mangin, Stéphane},
title = {Deterministic Ultra-Fast All-Optical Switching in Gd free Ferrimagnetic Spin Valve Structures},
journal = {Advanced Functional Materials},
volume = {35},
number = {52},
pages = {e05423},
doi = {https://doi.org/10.1002/adfm.202505423},
url = {https://advanced.onlinelibrary.wiley.com/doi/abs/10.1002/adfm.202505423},
year = {2025}
}

@article{Kim2022b,
author = {Kim, Hyun-Joong and Moon, Kyoung-Woong and Tran, Bao Xuan and Yoon, Seongsoo and Kim, Changsoo and Yang, Seungmo and Ha, Jae-Hyun and An, Kyongmo and Ju, Tae-Seong and Hong, Jung-Il and Hwang, Chanyong},
title = {Field-Free Switching of Magnetization by Tilting the Perpendicular Magnetic Anisotropy of Gd/Co Multilayers},
journal = {Advanced Functional Materials},
volume = {32},
number = {26},
pages = {2112561},
doi = {https://doi.org/10.1002/adfm.202112561},
year = {2022}
}

@article{Liu2024,
  title = {Manipulation of spin-orbit-torque efficiency via designing gradient structure in ferrimagnetic $\mathrm{Ta}/\mathrm{Tb}$-$\mathrm{Co}/\mathrm{Pt}$ multilayers},
  author = {Liu, Long and Wu, Jinxiang and Ye, Zhixing and Zhao, Xiaotian and Liu, Wei and Zhang, Zhidong},
  journal = {Phys. Rev. Appl.},
  volume = {21},
  issue = {4},
  pages = {044013},
  numpages = {13},
  year = {2024},
  month = {Apr},
  publisher = {American Physical Society},
  doi = {10.1103/PhysRevApplied.21.044013},
  url = {https://link.aps.org/doi/10.1103/PhysRevApplied.21.044013}
}

@article{Alebrand2012,
    author = {Alebrand, Sabine and Gottwald, Matthias and Hehn, Michel and Steil, Daniel and Cinchetti, Mirko and Lacour, Daniel and Fullerton, Eric E. and Aeschlimann, Martin and Mangin, StÃ©phane},
    title = "{Light-induced magnetization reversal of high-anisotropy TbCo alloy films}",
    journal = {Applied Physics Letters},
    volume = {101},
    number = {16},
    pages = {162408},
    year = {2012},
    month = {10},
}

@article{Wong2019,
  title = {Enhanced Spin-Orbit Torques in Rare-Earth $\mathrm{Pt}/[\mathrm{Co}/\mathrm{Ni}{]}_{2}/\mathrm{Co}/\mathrm{Tb}$ Systems},
  author = {Wong, Q.Y. and Murapaka, C. and Law, W.C. and Gan, W.L. and Lim, G.J. and Lew, W.S.},
  journal = {Phys. Rev. Appl.},
  volume = {11},
  issue = {2},
  pages = {024057},
  numpages = {10},
  year = {2019},
  month = {Feb},
  publisher = {American Physical Society},
  doi = {10.1103/PhysRevApplied.11.024057},
  url = {https://link.aps.org/doi/10.1103/PhysRevApplied.11.024057}
}

@article{Wu2020,
doi = {10.35848/1347-4065/aba793},
url = {https://doi.org/10.35848/1347-4065/aba793},
year = {2020},
month = {aug},
publisher = {IOP Publishing},
volume = {59},
number = {8},
pages = {080905},
author = {Wu, Yong and Chen, Jikun and Meng, Kangkang and Li, Zhipeng and Xu, Xiaoguang and Miao, Jun and Jiang, Yong},
title = {Perpendicular magnetic anisotropy and magnetization process of ferrimagnetic CoFeB/Tb multilayer films},
journal = {Japanese Journal of Applied Physics}}

@Article{Hu2024,
  author   = {Hu, Chen-Yu and Chen, Wei-De and Liu, Yan-Ting and Huang, Chao-Chung and Pai, Chi-Feng},
  title    = {The central role of tilted anisotropy for field-free spin-orbit torque switching of perpendicular magnetization},
  doi      = {10.1038/s41427-023-00521-9},
  issn     = {1884-4057},
  number   = {1},
  pages    = {1},
  url      = {https://doi.org/10.1038/s41427-023-00521-9},
  volume   = {16},
  journal  = {NPG Asia Materials},
  refid    = {Hu2024},
  year     = {2024},
}

@article{Mondal2023,
title = {Single-shot switching in Tb/Co-multilayer based nanoscale magnetic tunnel junctions},
journal = {Journal of Magnetism and Magnetic Materials},
volume = {581},
pages = {170960},
year = {2023},
issn = {0304-8853},
author = {Sucheta Mondal and Debanjan Polley and Akshay Pattabi and Jyotirmoy Chatterjee and David Salomoni and Luis Aviles-Felix and Aurélien Olivier and Miguel Rubio-Roy and Bernard Diény and Liliana Daniela Buda Prejbeanu and Ricardo Sousa and Ioan Lucian Prejbeanu and Jeffrey Bokor}
}

@article{Olivier2020,
year = {2020},
month = {jul},
publisher = {IOP Publishing},
volume = {31},
number = {42},
pages = {425302},
author = {A Olivier and L Avilés-Félix and A Chavent and L Álvaro-Gómez and M Rubio-Roy and S Auffret and L Vila and B Dieny and R C Sousa and I L Prejbeanu},
title = {Indium Tin Oxide optical access for magnetic tunnel junctions in hybrid spintronic–photonic circuits},
journal = {Nanotechnology}
}

@article{Mishra2023,
  title = {Dynamics of all-optical single-shot switching of magnetization in Tb/Co multilayers},
  author = {Mishra, K. and Blank, T. G. H. and Davies, C. S. and Avil\'es-F\'elix, L. and Salomoni, D. and Buda-Prejbeanu, L. D. and Sousa, R. C. and Prejbeanu, I. L. and Koopmans, B. and Rasing, Th. and Kimel, A. V. and Kirilyuk, A.},
  journal = {Phys. Rev. Res.},
  volume = {5},
  issue = {2},
  pages = {023163},
  numpages = {13},
  year = {2023},
  month = {Jun},
  publisher = {American Physical Society}
}
\end{document}